\documentclass[final,3p,times]{elsarticle} 

\usepackage{amsmath}

\usepackage{amssymb}
\usepackage{amsthm}

\newtheorem{lemma}{Lemma}[section]

\newtheorem{example}{Example}[section]
\newtheorem{definition}{Definition}[section]
\newtheorem{corollary}{Corollary}[section]
\newtheorem{theorem}{Theorem}[section]

\usepackage[ruled,linesnumbered]{algorithm2e}
\usepackage{tikz}
\usepackage{caption, subcaption}

\usepackage[labelfont=bf,figurename=Fig. ,labelsep=period,singlelinecheck=true]{caption}

\usepackage{mathtools} 

\DeclarePairedDelimiter{\ceil}{\lceil}{\rceil}

\usepackage{lineno}

\journal{Theoretical Computer Science}

\begin{document}

\begin{frontmatter}



\title{Packing squares independently}


\author[inst1]{Wei Wu}
\affiliation[inst1]{
            organization={Graduate School of Integrated Science and Technology, Shizuoka University},
            addressline={3-5-1 Johoku, Naka-ku},
            city={Hamamatsu},
            postcode={432-8561},
            state={Shizuoka},
            country={Japan}}
            
\author[inst2]{Hiroki Numaguchi}
\affiliation[inst2]{
            organization={Department of Applied Mathematics, Tokyo University of Science},
            addressline={1-3 Kagurazaka}, 
            city={Shinjuku-ku},
            postcode={162-8601}, 
            state={Tokyo},
            country={Japan}}
\author[inst2]{Yannan Hu}
\author[inst3]{Mutsunori Yagiura}
\affiliation[inst3]{
            organization={Graduate School of Fnformatics, Nagoya University},
            addressline={Furo-cho, Chikusa-ku}, 
            city={Nagoya},
            postcode={162-8601}, 
            state={Aichi},
            country={Japan}}

\begin{abstract}
Given a set of squares and a strip of bounded width and infinite height,
we consider a square strip packaging problem, which we call the square independent packing problem (SIPP),
to minimize the strip height so that all the squares are packed into independent cells separated by horizontal and vertical partitions.
For the SIPP, we first investigate efficient solution representations
and propose a compact representation that reduces the search space from $\Omega(n!)$ to $\mathcal{O}(2^n)$,
with $n$ the number of given squares, while guaranteeing that there exists a solution representation that corresponds to an optimal solution.
Based on the solution representation,
we show that the problem is $\mathcal{NP}$-hard, 
and then we propose a fully polynomial-time approximation scheme (FPTAS) to solve it.
We also propose three mathematical programming formulations based on different solution representations and confirm the performance of these algorithms through computational experiments.
Finally, we discuss several extensions that are relevant to practical applications.
\end{abstract}




\end{frontmatter}



\section{Introduction}
\label{sec:introduction}
This paper considers the square independent packing problem (SIPP),
which is a two-dimensional square strip packing problem to minimize the strip height with an additional constraint that requires all squares to be separated from each other by horizontal and vertical partitions, each going through one edge to the opposite side of the strip.

In order to meet the needs of practical applications, a large number of researchers have investigated various problems and made contributions in the field of cutting and packing.
Dyckhoff~\cite{dyckhoff1990typology} proposed a typology of cutting and packing problems, which was then improved by W{\"a}scher~et~al.~\cite{wascher2007improved}.
For the two-dimensional strip packing problem (see surveys~\cite{junior2022rectangular,lodi2002two}),
a number of models~\cite{bezerra2020models} and algorithms~\cite{bettinelli2008branch} have been extensively studied,
and several variations involving different real-life practical constraints have been considered as well.
Two relevant variations as relaxations of the SIPP are the strip packing problem with guillotine cut constraints~\cite{junior2022rectangular}, and the strip level packing problem~\cite{lodi2004models} in which items have to be packed ``in levels.''
However, even though the SIPP has many practical applications in areas such as logistics and industrial machinery design,
to the best of our knowledge, this is the first study that considers a strip packing problem in which each item should be packed into a separated cell.

In this paper, we first propose three types of solution representations, one of which that we call the row-column (RC) representation, is a compact representation that allows a solution with $p$ rows and $q$ columns to be represented by a binary vector of length $p+q$, instead of a mapping from $\{1,2,\ldots,n\}$ to $pq$ cells, reducing the search space from $\Omega(n!)$ to $\mathcal{O}(2^n)$,
with $n$ the number of given squares, while guaranteeing that there exists a solution representation that corresponds to an optimal solution.
To show the $\mathcal{NP}$-hardness of the SIPP, a simple reduction from an $\mathcal{NP}$-complete problem seems to be difficult due to the property that only the largest square in each row and column contributes to the overall size.
Using the RC representation, we show that the SIPP is $\mathcal{NP}$-hard.
To solve the SIPP, we propose three mathematical formulations based on different solution representations, and we evaluate them through computational experiments.
Although the formulation based on the RC representation is an integer quadratically constrained programming model, whereas the other two are mixed integer linear programming models, it performs significantly better to the other two formulations in terms of CPU time required because of the smaller number of variables used.
Furthermore, we design a fully polynomial-time approximation scheme (FPTAS) for the SIPP with a time complexity of $\mathcal{O}(n^{3.5}/\epsilon)$.

Finally, we discuss several extensions and provide interesting theoretical results on these extensions, including an extension that considers the $k$-dimensional SIPP ($k$SIPP).
For the $k$SIPP, we design a dynamic programming method with a time complexity of $O(kn^{h(k)}\prod_{i=1}^{k-1}b_i)$, where $h(k)=\sum_{i=1}^k \frac{1}{i}$ and $b_i$ is the capacity for each dimension $i=1,2,\dots,k-1$.
However, for a $k$SIPP with $k\ge 3$, we show that there is no FPTAS unless $\mathcal{P}=\mathcal{NP}$.


\section{Problem description}
\label{sec:problem-description}
We are given a strip of width $b$ and a set of squares $N=\{1,2, \dots,n\}$, each square $i$ with a side length $l_i$.
The strip can be separated into cells by placing vertical (resp. horizontal) partitions from top to bottom (resp. left side to right side) of the strip.
A vertical (resp. horizontal) partition can be placed at any horizontal (resp. vertical) position of the strip (i.e., at any $x$-coordinate in range $[0, b]$
(resp. nonnegative $y$-coordinate), considering the bottom-left corner of the strip as the origin, and the bottom (resp. left) edge of the strip as the $x$-axis (resp. $y$-axis)).
The \textsc{Square Independent Packing Problem} (SIPP) aims to minimize the height of the strip so that all the squares are packed in independent cells separated by partitions in such a way that every cell contains at most one square.

We assume that the thickness of a partition is 0, and all the input values are positive integers.
(The case involving positive-thickness partitions is discussed in Section~\ref{sec:extension}.)
For the remainder of this paper, we also assume that the squares are sorted and numbered in non-increasing order of their side lengths, that is,
\begin{align}\label{assp:sprted}
    l_1\ge l_2 \ge \dots \ge l_n.
\end{align}

\section{Solution representations}\label{sec:solution-represntation}
\begin{figure}\centering
    \begin{subfigure}[b]{0.3\textwidth}\centering\scriptsize
        \begin{tikzpicture}[scale=0.65]
            \draw[line width = 2 pt] (0  pt, 0  pt) -- (186  pt, 0  pt);
            \draw[line width = 2 pt] (0  pt, 47  pt) -- (186  pt, 47  pt);
            \draw[line width = 2 pt] (0  pt, 82  pt) -- (186  pt, 82  pt);
            \draw[line width = 2 pt] (0  pt, 144  pt) -- (186  pt, 144  pt);
            \draw[line width = 2 pt] (0  pt, 0  pt) -- (0  pt, 144  pt);
            \draw[line width = 2 pt] (41  pt, 0  pt) -- (41  pt, 144  pt);
            \draw[line width = 2 pt] (103  pt, 0  pt) -- (103  pt, 144  pt);
            \draw[line width = 2 pt] (186  pt, 0  pt) -- (186  pt, 144  pt);
            \draw[draw=black, fill=black!8] (1  pt, 1  pt) rectangle ++ (39  pt, 39  pt)  node[pos=.5] {3};
            \draw[draw=black, fill=black!8] (42  pt, 1  pt) rectangle ++ (15  pt, 15  pt)  node[pos=.5] {7};
            \draw[draw=black, fill=black!8] (104  pt, 1  pt) rectangle ++ (45  pt, 45  pt)  node[pos=.5] {2};
            \draw[draw=black, fill=black!8] (1  pt, 48  pt) rectangle ++ (33  pt, 33  pt)  node[pos=.5] {5};
            \draw[draw=black, fill=black!8] (42  pt, 48  pt) rectangle ++ (24  pt, 24  pt)  node[pos=.5] {6};
            \draw[draw=black, fill=black!8] (1  pt, 83  pt) rectangle ++ (39  pt, 39  pt)  node[pos=.5] {4};
            \draw[draw=black, fill=black!8] (42  pt, 83  pt) rectangle ++ (60  pt, 60  pt)  node[pos=.5] {1};
            \draw[draw=black, fill=black!8] (104  pt, 83  pt) rectangle ++ (9  pt, 9  pt)  node[pos=.5] {8};
        \end{tikzpicture}
        \caption{A solution represented by a matrix}
        \label{fig:eg-matrix}
    \end{subfigure}
    \hfill
    \begin{subfigure}[b]{0.3\textwidth}\centering\scriptsize
        \begin{tikzpicture}[scale=0.65]
            \draw[line width = 2 pt] (0  pt, 0  pt) -- (186  pt, 0  pt);
            \draw[line width = 2 pt] (0  pt, 62  pt) -- (186  pt, 62  pt);
            \draw[line width = 2 pt] (0  pt, 103  pt) -- (186  pt, 103  pt);
            \draw[line width = 2 pt] (0  pt, 138  pt) -- (186  pt, 138  pt);
            \draw[line width = 2 pt] (0  pt, 0  pt) -- (0  pt, 138  pt);
            \draw[line width = 2 pt] (62  pt, 0  pt) -- (62  pt, 138  pt);
            \draw[line width = 2 pt] (109  pt, 0  pt) -- (109  pt, 138  pt);
            \draw[line width = 2 pt] (186  pt, 0  pt) -- (186  pt, 138  pt);
            \draw[draw=black, fill=black!8] (1  pt, 1  pt) rectangle ++ (60  pt, 60  pt)  node[pos=.5] {1};
            \draw[draw=black, fill=black!8] (63  pt, 1  pt) rectangle ++ (45  pt, 45  pt)  node[pos=.5] {2};
            \draw[draw=black, fill=black!8] (110  pt, 1  pt) rectangle ++ (39  pt, 39  pt)  node[pos=.5] {3};
            \draw[draw=black, fill=black!8] (1  pt, 63  pt) rectangle ++ (39  pt, 39  pt)  node[pos=.5] {4};
            \draw[draw=black, fill=black!8] (63  pt, 63  pt) rectangle ++ (24  pt, 24  pt)  node[pos=.5] {6};
            \draw[draw=black, fill=black!8] (110  pt, 63  pt) rectangle ++ (9  pt, 9  pt)  node[pos=.5] {8};
            \draw[draw=black, fill=black!8] (1  pt, 104  pt) rectangle ++ (33  pt, 33  pt)  node[pos=.5] {5};
            \draw[draw=black, fill=black!8] (63  pt, 104  pt) rectangle ++ (15  pt, 15  pt)  node[pos=.5] {7};
        \end{tikzpicture}
        \caption{A sorted layout transformed from the solution in \figurename~\ref{fig:eg-matrix}}
        \label{fig:eg-sorted}
    \end{subfigure}
    \hfill
    \begin{subfigure}[b]{0.3\textwidth}\centering\scriptsize
        \begin{tikzpicture}[scale=0.65]
            \draw[line width = 2 pt] (0  pt, 0  pt) -- (167  pt, 0  pt);
            \draw[line width = 2 pt] (0  pt, 62  pt) -- (167  pt, 62  pt);
            \draw[line width = 2 pt] (0  pt, 103  pt) -- (167  pt, 103  pt);
            \draw[line width = 2 pt] (0  pt, 0  pt) -- (0  pt, 103  pt);
            \draw[line width = 2 pt] (62  pt, 0  pt) -- (62  pt, 103  pt);
            \draw[line width = 2 pt] (109  pt, 0  pt) -- (109  pt, 103  pt);
            \draw[line width = 2 pt] (150  pt, 0  pt) -- (150  pt, 103  pt);
            \draw[line width = 2 pt] (167  pt, 0  pt) -- (167  pt, 103  pt);
            \draw[draw=black, fill=black!8] (1  pt, 1  pt) rectangle ++ (60  pt, 60  pt)  node[pos=.5] {1};
            \draw[draw=black, fill=black!8] (63  pt, 1  pt) rectangle ++ (45  pt, 45  pt)  node[pos=.5] {2};
            \draw[draw=black, fill=black!8] (110  pt, 1  pt) rectangle ++ (39  pt, 39  pt)  node[pos=.5] {3};
            \draw[draw=black, fill=black!8] (151  pt, 1  pt) rectangle ++ (15  pt, 15  pt)  node[pos=.5] {7};
            \draw[draw=black, fill=black!8] (1  pt, 63  pt) rectangle ++ (39  pt, 39  pt)  node[pos=.5] {4};
            \draw[draw=black, fill=black!8] (63  pt, 63  pt) rectangle ++ (33  pt, 33  pt)  node[pos=.5] {5};
            \draw[draw=black, fill=black!8] (110  pt, 63  pt) rectangle ++ (24  pt, 24  pt)  node[pos=.5] {6};
            \draw[draw=black, fill=black!8] (151  pt, 63  pt) rectangle ++ (9  pt, 9  pt)  node[pos=.5] {8};
        \end{tikzpicture}
        \caption{An RC layout with $\boldsymbol{\alpha}=(\mathrm{AC},\mathrm{AC},\mathrm{AR},\mathrm{AC})$}
        \label{fig:eg-rc}
     \end{subfigure}
        \caption{Three solutions for the instance in Example~\ref{exp:ins}}
        \label{fig:three graphs}
\end{figure}
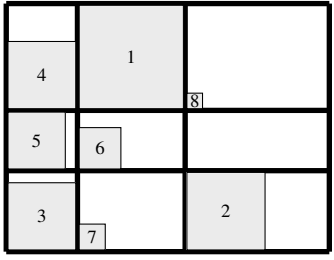
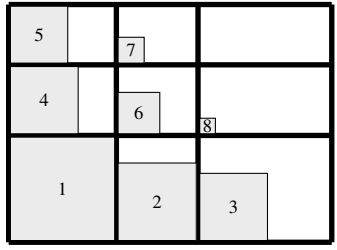
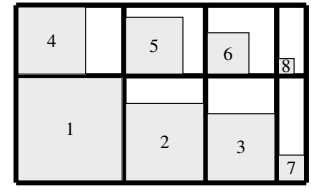

A natural solution representation to the SIPP is to use a matrix $\boldsymbol{A}=(a_{ij})\in \mathbb{N}^{p\times q}$, where $a_{ij}$ denotes the square placed in the $i$th row from the bottom and the $j$th column from the left.
Such a matrix $\boldsymbol{A}$ must contain values $1,2,\dots,n$.
For convenience, we assume that dummy squares $n+1,n+2, \dots, pq$ of side length $0$ are contained in matrix $\boldsymbol{A}$ if $pq>n$.
In this paper, a matrix $\boldsymbol{A}=(a_{ij})\in \mathbb{N}^{p\times q}$ is called a \emph{layout} if it contains all the values from $1$ to $pq$.
Any feasible solution to the SIPP can be represented by a layout $\boldsymbol{A}=(a_{ij})\in \mathbb{N}^{p\times q}$ with $pq\ge n$.
\begin{example}\label{exp:ins}
Consider an SIPP instance with
\begin{align*}
    n=8, \quad\boldsymbol{l}=(l_1,l_2,\dots,l_8) = (20, 15, 13, 13, 11, 8, 5, 3), \quad b=60.
\end{align*}
\end{example}
For the instance in Example~\ref{exp:ins},
\figurename~\ref{fig:eg-matrix} shows a solution, which can be represented by
\begin{align}\label{exp:layout}
    \boldsymbol{A}^{(\mathrm{a})} = 
\begin{pmatrix}
a^{(\mathrm{a})}_{11} & a^{(\mathrm{a})}_{12} & a^{(\mathrm{a})}_{13} \\
a^{(\mathrm{a})}_{21} & a^{(\mathrm{a})}_{22} & a^{(\mathrm{a})}_{23} \\
a^{(\mathrm{a})}_{31} & a^{(\mathrm{a})}_{32} & a^{(\mathrm{a})}_{33}
\end{pmatrix}= 
\begin{pmatrix}
3 & 7 & 2 \\
5 & 6 & 9 \\
4 & 1 & 8
\end{pmatrix}
\end{align}
with a dummy square 9.

We use $W(\boldsymbol{A})$ and $H(\boldsymbol{A})$ to denote the width and height of layout $\boldsymbol{A}$, respectively.
For solution $\boldsymbol{A}^{(\mathrm{a})}$ to the instance in Example~\ref{exp:ins}, the width and height are calculated as
\begin{align*}
    W(\boldsymbol{A}^{(\mathrm{a})})&=\sum_{j=1}^{q} \max_{i\in\{1,\ldots,p\}}l_{a^{(\mathrm{a})}_{ij}}
    &H(\boldsymbol{A}^{(\mathrm{a})})&=\sum_{i=1}^{p} \max_{j\in\{1,\ldots,q\}}l_{a^{(\mathrm{a})}_{ij}}\\
    &=l_3+l_1+l_2=48,&
    &=l_2+l_5+l_1=46.
\end{align*}
Because the layout width $W(\boldsymbol{A}^{(\mathrm{a})})$ does not exceed the strip width $b=60$, $\boldsymbol{A}^{(\mathrm{a})}$ represents a feasible solution whose objective function value (height) is 46.

\subsection{Sorted layout}
\begin{definition}[sorted layout]\label{def:SL}
    A layout $\boldsymbol{A}$ is a sorted layout, if the values in each row and each column are sorted in ascending order.
\end{definition}
From the assumption in~\eqref{assp:sprted}, 
a sorted layout represents a solution for which
the squares in each row (resp. column) are sorted from left to right (resp. bottom to top) in non-ascending order of their side lengths.
For example, 
\begin{align*}
    \boldsymbol{A}^{(\mathrm{b})} = 
\begin{pmatrix}
1 & 2 & 3 \\
4 & 6 & 8 \\
5 & 7 & 9
\end{pmatrix}
\end{align*}
is a sorted layout, while $\boldsymbol{A}^{(\mathrm{a})}$ in \eqref{exp:layout} is not.
\figurename~\ref{fig:eg-sorted} visualizes solution $\boldsymbol{A}^{(\mathrm{b})}$
for the instance in Example~\ref{exp:ins}.
For a sorted layout, the width (resp. the height) is equal to the total side lengths of the squares in the first row (resp. the first column).
For the instance in Example~\ref{exp:ins}, we have $W(\boldsymbol{A}^{(\mathrm{b})})=l_1+l_2+l_3=48$ and $H(\boldsymbol{A}^{(\mathrm{b})})=l_1+l_4+l_5=44$.

\begin{lemma}
\label{lem:ext-SL}
    An arbitrary layout can be transformed to a sorted layout without increasing its width and height.
\end{lemma}
\begin{proof}
Given a layout $\boldsymbol{A'}=(a'_{ij})\in \mathbb{N}^{p\times q}$ for the SIPP,
we show there exists an algorithm that transforms $\boldsymbol{A'}$ to a sorted layout $\boldsymbol{A}^\mathrm{sort}=(a^\mathrm{sort}_{ij})\in \mathbb{N}^{p\times q}$ without increasing the width and height.
The pseudo-code of the transformation is shown in Algorithm~\ref{algo:sorted}.
\begin{algorithm}[ht]
    \caption{Transform a layout $\boldsymbol{A'}$ to a sorted layout $\boldsymbol{A}^\mathrm{sort}$}\label{algo:sorted}
    Let $\boldsymbol{A}^\mathrm{sort}\in \mathbb{N}^{p\times q}$ be an empty matrix\;
    \For{$k\gets 1$ \KwTo $p$}{
    Find the $i'$th row that contains the minimum value (the largest square) in $\boldsymbol{A'}$\;\label{step:sorted-findrowmin}
    \lIf{$i'\ne 1$}{swap the first row and the $i'$th row of matrix $\boldsymbol{A'}$}\label{step:sorted-swaprow}
    \lForEach{\rm{column in matrix} $\boldsymbol{A'}$}{swap the first value and the minimum value in the column}\label{step:sorted-swapcolitem}
    Sort the columns of $\boldsymbol{A'}$ in increasing order by the first value of the column\;\label{step:sorted-sortcol}
    Copy the first row of $\boldsymbol{A'}$ to the $k$th row of $\boldsymbol{A}^\mathrm{sort}$\;\label{step:sorted-copy} 
    Remove the first row from $\boldsymbol{A'}$, that is, $\boldsymbol{A'}$ becomes a matrix with $(p-k)$ rows and $q$ columns\;\label{step:sorted-remove} 
   }
    \Return $\boldsymbol{A}^\mathrm{sort}$\;
\end{algorithm}
In Step~\ref{step:sorted-findrowmin}--\ref{step:sorted-swaprow}, swapping two rows does not change the height and width of $\boldsymbol{A'}$.
Because in Step~\ref{step:sorted-swaprow} the largest square in each column is moved to the first row, swap operations in Step~\ref{step:sorted-swapcolitem} also do not increase the height of the corresponding rows.
For the first loop ($k= 1$), before Step~\ref{step:sorted-sortcol}, we have
\begin{align}\label{condition:base}
    &a'_{1j} < \min_{i=2}^{p}\{a'_{ij}\}&\forall j\in\{1,2,\dots,q\}.
\end{align}
Step~\ref{step:sorted-sortcol} reorders the columns, but does not change the values in each columns, so after Step~\ref{step:sorted-sortcol} \eqref{condition:base} still holds true and
\begin{align*}
    &a'_{11} < a'_{12} < \dots < a'_{1q}
\end{align*}
becomes to be satisfied.
In Step~\ref{step:sorted-copy} and \ref{step:sorted-remove}, we copy the first row of $\boldsymbol{A'}$ to $\boldsymbol{A}^\mathrm{sort}$ and then remove the row from $\boldsymbol{A'}$.
Thus, at the end of the loop,
\begin{align}
\label{condition:col}
    &a^\mathrm{sort}_{kj} < \min_{i=1}^{p-k}\{a'_{ij}\}&\forall j\in\{1,2,\dots,q\}\\
\label{condition:row}
    &a^\mathrm{sort}_{k1} < a^\mathrm{sort}_{k2} < \dots < a^\mathrm{sort}_{kq}
\end{align}
hold for $k=1$.
For the case when $k= 2$, because \eqref{condition:col} and \eqref{condition:row} hold at the end of the first loop,
we have 
\begin{align*}
    &a^\mathrm{sort}_{(k-1),j} < \min_{i=1}^{p-k+1}\{a'_{ij}\} = a'_{1j}&\forall j\in\{1,2,\dots,q\}\\
    &a^\mathrm{sort}_{k-1,1} < a^\mathrm{sort}_{k-1,2} < \dots < a^\mathrm{sort}_{k-1,q}
\end{align*}
before Step~\ref{step:sorted-sortcol},
which implies that the sorting operation in Step~\ref{step:sorted-sortcol} will not change the width of $\boldsymbol{A}^\mathrm{sort}$ after we copy the first row of $\boldsymbol{A'}$ to $\boldsymbol{A}^\mathrm{sort}$.
Hence, inequalities~\eqref{condition:col} and~\eqref{condition:row} also hold true at the end of the second loop ($k=2$).
Because the same description is valid for the case when $k\ge 3$, from \eqref{condition:col} and~\eqref{condition:row} for each $k$, we can obtain a layout $\boldsymbol{A}^\mathrm{sort}$ in which
\begin{align*}
    &a^\mathrm{sort}_{1j} < a^\mathrm{sort}_{2j} < \dots < a^\mathrm{sort}_{pj}&\forall j\in\{1,2,\dots,q\}\\
    &a^\mathrm{sort}_{i1} < a^\mathrm{sort}_{i2} < \dots < a^\mathrm{sort}_{iq}&\forall i\in\{1,2,\dots,p\}
\end{align*}
hold.

Algorithm~\ref{algo:sorted} can transform any layout to a sorted layout with the same width and a non-increasing height.
For the instance in Example~\ref{exp:ins},
layout~$\boldsymbol{A}^{(\mathrm{b})}$ (\figurename~\ref{fig:eg-sorted}) is the sorted layout obtained by transforming~$\bar{\boldsymbol{A}}$ (\figurename~\ref{fig:eg-matrix}).
Both layouts have the same width and $\boldsymbol{A}^{(\mathrm{b})}$ is 2 units shorter than $\bar{\boldsymbol{A}}$ in height.
\end{proof}

\begin{corollary}\label{cly:opt-SL}
    There exists a sorted layout that is optimal to the SIPP.
\end{corollary}
\begin{proof}
    For any optimal layout,
    Lemma~\ref{lem:ext-SL} indicates that there always exists a sorted layout with the same or shorter width and height.
    Such a sorted layout must be optimal to the SIPP.
\end{proof}
Corollary~\ref{cly:opt-SL} indicates that we can restrict the search space to all sorted layouts when solving the SIPP.

We also show a property in Lemma~\ref{lem:sorted-next} that will be used in later proofs.
For a layout $\boldsymbol{A}=(a_{ij})\in \mathbb{N}^{p\times q}$, we use $\boldsymbol{A}_{p'q'}$ to denote the submatrix $(a_{ij})_{\substack{i=1,2,\dots,p'\\j=1,2,\dots,q'}}$ of $\boldsymbol{A}$.
\begin{lemma}\label{lem:sorted-next}
    For a sorted layout $\boldsymbol{A}=(a_{ij})\in \mathbb{N}^{p\times q}$, if its submatrix $\boldsymbol{A}_{p'q'}$ ($p'q'<pq$) is a sorted layout, then either $a_{1,q'+1}=p'q'+1$ or $a_{p'+1,1}=p'q'+1$ holds true.
\end{lemma}
\begin{proof}
    Because $\boldsymbol{A}_{p'q'}$ is a (sorted) layout with $p'$ rows and $q'$ columns, we know that it does not contain $p'q'+1$.
    From the fact that $\boldsymbol{A}$ is sorted, we can further be certain that value $p'q'+1$ is stored in $a_{1,q'+1}$ or $a_{p'+1,1}$.
\end{proof}

\subsection{Row-column layout and row-column sequence}
For a $p\times q$ layout, an \emph{add-row} (AR) operation is defined to be an operation that adds a new row $(pq+1,pq+2,\dots,pq+q)$ after row $p$.
Similarly, an \emph{add-column} (AC) operation adds a new column $(pq+1,pq+2,\dots,pq+p)^{\intercal}$ after column $q$.
For the layout $\boldsymbol{A}^{(\mathrm{a})}$ in \eqref{exp:layout},
\begin{align*}
\begin{pmatrix}
3 & 7 & 2 \\
5 & 6 & 9 \\
4 & 1 & 8 \\
10 & 11 & 12
\end{pmatrix}
\qquad\text{and}\qquad
\begin{pmatrix}
3 & 7 & 2 & 10 \\
5 & 6 & 9 & 11 \\
4 & 1 & 8 & 12 
\end{pmatrix}
\end{align*}
are the layouts obtained after an AR operation and AC operation, respectively.

In this paper, the $1\times 1$ layout (that contains value 1) is called the \emph{base layout}.

\begin{definition}[row-column layout]\label{def:RCL}
    A layout $\boldsymbol{A}$ is a row-column (RC) layout, if it can be created from the base layout by a sequence of the AR and AC operations.
\end{definition}
From Definition~\ref{def:RCL}, an RC layout of size $p\times q$ can be represented by a sequence of length $p+q-2$ with $p-1$ AR operations and $q-1$ AC operations, and vice versa.
In this paper, we use the following notation to represent such a sequence.
\begin{definition}[row-column sequence]\label{def:RCS}
    A sequence $\boldsymbol{\alpha}=(\alpha_1,\alpha_2,\dots,\alpha_u)$ is a row-column (RC) sequence, if $\alpha_i \in \{\mathrm{AR}, \mathrm{AC}\}$ for each $i=1,2,\dots,u$.
\end{definition}
For convenience, we use $W(\boldsymbol{\alpha})$ and $H(\boldsymbol{\alpha})$ to denote the width and height of the solution corresponding to an RC sequence $\boldsymbol{\alpha}$, respectively.
For the instance in Example~\ref{exp:ins},
\figurename~\ref{fig:eg-rc} visualizes a solution corresponding to an RC sequence $\boldsymbol{\alpha}=(\mathrm{AC},\mathrm{AC},\mathrm{AR},\mathrm{AC})$ with $W(\boldsymbol{\alpha})=53$ and $H(\boldsymbol{\alpha})=33$.
Because $W(\boldsymbol{\alpha})=53<b=60$, $\boldsymbol{\alpha}$ is a feasible solution to the SIPP.

From the definitions of the sorted layout (Definition~\ref{def:SL}) and the RC layout (Definition~\ref{def:RCL}),
RC layout is a special case of sorted layout.

For these layouts, we have the following lemma.
\begin{lemma}
\label{lem:ext-RC}
    A sorted layout can be transformed to an RC layout without increasing the width and height.
\end{lemma}
\begin{proof}
We show that there exists an algorithm that transforms a sorted layout $\boldsymbol{\tilde{A}}=(\tilde{a}_{ij})\in \mathbb{N}^{p\times q}$ to an RC layout without increasing the width and the height.
The transformation is shown in Algorithm~\ref{algo:RC}.
\begin{algorithm}[ht]
    \caption{Transform a sorted layout $\boldsymbol{\tilde{A}}=(\tilde{a}_{ij})\in \mathbb{N}^{p\times q}$ to an RC layout}\label{algo:RC}
    $p'\gets 1, q'\gets 1$\;
    \While{\rm{$p'<p$ \KwSty{or} $q'<q$}}{
    \If{$\tilde{a}_{1,q'+1}=p'q'+1$}{\label{step:rc-coladd-start}
        \For{$k\gets 2$ \KwTo $p'$}{
            \While{$\tilde{a}_{k,q'+1}> p'q'+k$}{
                Find the cell $\tilde{a}_{uv}$ that stores value $\tilde{a}_{k,q'+1}-1$\;
                Swap the values in $\tilde{a}_{uv}$ and $\tilde{a}_{k,q'+1}$\;
           }
       }
        $q'\gets q'+1$\;
   }\label{step:rc-coladd-end}
    \If{$\tilde{a}_{p'+1,1}=p'q'+1$}{\label{step:rc-rowadd-start}
        \For{$k\gets 2$ \KwTo $q'$}{
            \While{$\tilde{a}_{p'+1,k}> p'q'+k$}{
                Find the cell $\tilde{a}_{uv}$ that stores value $\tilde{a}_{p'+1,k}-1$\;
                Swap the values in $\tilde{a}_{uv}$ and $\tilde{a}_{p'+1,k}$\;
           }
       }
        $p'\gets p'+1$\;
   }\label{step:rc-rowadd-end}
   }
\end{algorithm}
We will show that during each step of Algorithm~\ref{algo:RC}, the following properties are always be satisfied:
\begin{enumerate}
    \item Layout $\boldsymbol{\tilde{A}}$ is a sorted layout.\label{prop:proof-sorted}
    \item The height and the width of $\boldsymbol{\tilde{A}}$ do not increase.\label{prop:proof-noinc}
    \item Layout $\boldsymbol{\tilde{A}}_{p'q'}$ is an RC layout.\label{prop:proof-RC}
\end{enumerate}
Because layout $\boldsymbol{\tilde{A}}$ before the transformation is a sorted layout, it must contain the base layout as its submatrix, which is also an RC layout.
That is, the initial $\boldsymbol{\tilde{A}}$ contains an RC layout $\boldsymbol{\tilde{A}}_{p'q'}$, where $p'=1$ and $q'=1$.
The idea in the transformation is to extend the submatrix $\boldsymbol{\tilde{A}}_{p'q'}$ by adding a row or adding a column so that it is still an RC layout.
More specifically, because both $\boldsymbol{\tilde{A}}$ and $\boldsymbol{\tilde{A}}_{11}$ are sorted layouts, from Lemma~\ref{lem:sorted-next} we know that either $\tilde{a}_{12}=2$ or $\tilde{a}_{21}=2$ holds.
If $\tilde{a}_{12}=2$, then $\boldsymbol{\tilde{A}}_{p'q'}$ is an RC layout with $p'=1$ and $q'=2$, that is, $q'$ can be extended from 1 to 2; otherwise, if $\tilde{a}_{21}=2$, $p'$ can be extended to 2.
Using Lemma~\ref{lem:sorted-next} again, we have $\tilde{a}_{1,q'+1}=p'q'+1$ or $\tilde{a}_{p'+1,1}=p'q'+1$.
If $\tilde{a}_{1,q'+1}=p'q'+1$ holds, we try to fulfill an AC operation to the current $\boldsymbol{\tilde{A}}_{p'q'}$ (Step~\ref{step:rc-coladd-start}--\ref{step:rc-coladd-end}).
We check $\tilde{a}_{k,q'+1}$ for $k=2,\dots,p'$.
If $\tilde{a}_{k,q'+1}> p'q'+k$ was found, we find a cell $\tilde{a}_{uv}$ that stores the value $\tilde{a}_{k,q'+1}-1$.
Note that $u\ne k$ and $v\ne q'+1$ hold because of property~\eqref{prop:proof-sorted} and property~\eqref{prop:proof-RC}.
Thus, we can swap the values in $\tilde{a}_{uv}$ and $\tilde{a}_{k,q'+1}$ and then property~\eqref{prop:proof-sorted} still holds true.
Because $\tilde{a}_{uv}$ gets a value 1 larger than before (a non-larger square), property~\eqref{prop:proof-noinc} still holds after the swap operation.
It implies that property~\eqref{prop:proof-sorted} and property~\eqref{prop:proof-noinc} always hold true when iterating such swaps for $\tilde{a}_{k,q'+1}$ until $\tilde{a}_{k,q'+1}= p'q'+k$ is satisfied.
All these processes for each $k$ from 2 to $q'$ lead to an AC operation to the current RC layout $\boldsymbol{\tilde{A}}_{p'q'}$.
Otherwise, if $\tilde{a}_{p'+1,1}=p'q'+1$, an AR operation (Step~\ref{step:rc-rowadd-start}--\ref{step:rc-rowadd-end}) to extend $\boldsymbol{\tilde{A}}_{p'q'}$ is available.
Iterating such extensions on $\boldsymbol{\tilde{A}}_{p'q'}$ leads to a $p\times q$ RC layout without increasing the height and weight.
\end{proof}

\begin{corollary}\label{cly:opt-RCL}
    There exists an RC layout (RC sequence) optimal to the SIPP.
\end{corollary}
\begin{proof}
    From Lemma~\ref{lem:ext-RC}, each sorted layout can be transformed to an RC layout without increasing the height and weight.
Combining with the results in Corollary~\ref{cly:opt-SL} and the definition of the RC sequence, clearly, such an optimal RC layout exists.
\end{proof}
Corollary~\ref{cly:opt-RCL} indicates that we can solve the SIPP by further focusing on all the valid RC sequences.

\section{Complexity}\label{sec:complexity}
We begin by introducing a property of the sorted layout, which we use to demonstrate the complexity of the SIPP.
For a solution represented by a sorted layout with $p$ rows and $q$ columns,
we say that a value (square) is a \emph{bottleneck} value (square) if it is stored in the first row or column.
We use $v_k$ to denote the $k$th $(1\le k\le p+q-1)$ bottleneck value from $1$ to $pq$.
For the example shown in \figurename~\ref{fig:eg-rc},
the squares $1,2,3,4,7$ are bottleneck squares, and
\begin{align*}
    v_1=1,v_2=2,v_3=3,v_4=4,v_5=7.
\end{align*}

\begin{lemma}\label{lem:bottleneck}
    For any sorted layout $\boldsymbol{A}^\mathrm{sort}=(a^\mathrm{sort}_{ij})\in \mathbb{N}^{p\times q}$, $v_k\le \omega(k)$ holds for $k=1,2,\dots,p+q-1$, where
    \begin{align}\label{form:omega}
    \omega(k)=
        \begin{cases}
        \frac{(k+1)(k-1)}{4}+1 &\text{if $k$ is odd,}\\
        \frac{k^2}{4}+1 &\text{if $k$ is even.} 
        \end{cases}
    \end{align}
\end{lemma}
\begin{proof}
Because $\boldsymbol{A}^\mathrm{sort}$ is a sorted layout, the first bottleneck value $v_1$ is value 1.
Thus, for $k=1$, $v_k=1\le \omega(k)=1$ is correct.
Again, because $\boldsymbol{A}^\mathrm{sort}$ is a sorted layout,
for $k\ge 2$ we can assume that the first, second, $\dots, (k-1)$th bottleneck values are stored in $a^\mathrm{sort}_{11}, a^\mathrm{sort}_{12}, \dots, a^\mathrm{sort}_{1q'}$ and $a^\mathrm{sort}_{21}, a^\mathrm{sort}_{31}, \dots, a^\mathrm{sort}_{p',1}$, where
\begin{align}\label{proof:pqk}
    q'+p'=k.
\end{align}
The $k$th bottleneck value $v_k$ must be placed in $a^\mathrm{sort}_{p'+1,1}$ or $a^\mathrm{sort}_{1,q'+1}$, and $v_k=\min\left\{a^\mathrm{sort}_{p'+1,1},a^\mathrm{sort}_{1,q'+1}\right\}$ holds, which implies that
\begin{align*}
    &v_k\le a^\mathrm{sort}_{p'+1,1}\le a^\mathrm{sort}_{ij} &\forall i\in\{p'+1,p'+2,\dots,p\},\forall j\in \{1,2,\dots,q\}\\
    &v_k\le a^\mathrm{sort}_{1,q'+1}\le a^\mathrm{sort}_{ij} &\forall i\in\{1,2,\dots,p\},\forall j\in \{q'+1,q'+2,\dots,q\}.
\end{align*}
Thus, $v_k\le p'q'+1$ holds, because there exist at most $p'q'$ values that are less than $v_k$.
From~\eqref{proof:pqk} and the fact that both $q'$ and $p'$ are positive integers, we can further be certain that 
\begin{align*}
    v_k\le p'q'+1\le 
    \begin{cases}
    \frac{(k+1)(k-1)}{4}+1 &\text{if $k$ is odd,}\\
    \frac{k^2}{4}+1 &\text{otherwise.} 
    \end{cases}
\end{align*}
\end{proof}
From~\eqref{form:omega}, we can obtain the following equation:
\begin{align}\label{eq:bottleneck}
\sum_{j=1}^{2m+1} l_{\omega(j)}= l_{\omega(1)}+\sum_{i=1}^{m}l_{\omega(2i+1)}+\sum_{i=1}^{m}l_{\omega(2i)}= l_1+\sum_{i=1}^{m}l_{i(i+1)+1}+\sum_{i=1}^{m}l_{i^2+1}.
\end{align}


To show the complexity of the SIPP, we introduce the \textsc{Partition Problem} (PP), which is known to be $\mathcal{NP}$-complete~\cite{garey1979computers}.
Given $m$ positive integer numbers $s_1,s_2,\dots,s_m$, the PP asks if there exists a subset $S\subseteq M=\{1,2,\dots,m\}$ such that $\sum_{i\in S}s_i =\sum_{i\in M\setminus S}s_i$.
For convenience, we define $\beta$ as $\beta=\sum_{i\in M}s_i>0$.

\begin{theorem}\label{thm:sipp-complexity}
    The SIPP is $\mathcal{NP}$-hard.
\end{theorem}
\begin{proof}
We show the PP is reducible to the decision version of the SIPP (D-SIPP), which asks if there exists a solution such that the height of the strip is less than or equal to an input value $\lambda$.
Given an instance of the PP, we consider a D-SIPP instance as follows:
\begin{align}\label{proof:transfer-start}
    &n=(m+1)^2; \quad b=\lambda=\frac{(m+1)(m+2)}{2}\beta+\frac{1}{2}\beta; \quad l_1=(m+1)\beta,\\
    \label{proof:transfer-end}
    &\begin{cases}
    l_j = (m-i+1)\beta+s_i &\text{if $j=i^2+1,i^2+2,\dots,i(i+1)$}\\
    l_j = (m-i+1)\beta &\text{if $j=i(i+1)+1,i(i+1)+2,\dots,(i+1)^2$}\\
    \end{cases}
    &\forall i\in M.
\end{align}
Note that both the instance size of the D-SIPP and the transformation time is bounded by polynomials in the input size $m$ of the PP,
and the assumption in~\eqref{assp:sprted} holds true for this D-SIPP instance.
For example, when $m=3$, the D-SIPP instance becomes
\begin{align*}
    &n=16; \quad b=\lambda=\frac{21}{2}\beta; \quad l_1 = 4\beta,\\
    &l_2 = 3\beta+s_1, &&l_3 = l_4 = 3\beta,&\leftarrow i=1\\
    &l_5 = l_6 = 2\beta+s_2, &&l_7 = l_8 = l_9 = 2\beta,&\leftarrow i=2\\
    &l_{10} = l_{11} = l_{12} = \beta+s_3, &&l_{13} = l_{14} = l_{15} = l_{16} = \beta.&\leftarrow i=3
\end{align*}

If the answer to the PP is `yes', then there exists a subset $S'\subseteq M$ such that
\begin{align}\label{proof:ass-part}
    \sum_{i\in S'} s_i = \sum_{i\in M\setminus S'} s_i = \frac{1}{2}\beta.
\end{align}
In this case, an RC sequence $\boldsymbol{\alpha}'$ of length $2m$, where
\begin{align*}
&\begin{cases}
\alpha'_{2i} = \mathrm{AR}, \alpha'_{2i-1} = \mathrm{AC} &\text{if $i\in S'$}\\
\alpha'_{2i} = \mathrm{AC}, \alpha'_{2i-1} = \mathrm{AR} &\text{if $i\notin S'$}
\end{cases}&&\forall i\in M,
\end{align*}
leads to a solution to the D-SIPP with the following width and height:
\begin{align*}
    W(\boldsymbol{\alpha}') &= l_1 + \sum_{i\in S'} l_{i^2+1} + \sum_{i\in M\setminus S'} l_{i(i+1)+1}=(m+1)\beta+m\beta+\dots+\beta + \sum_{i\in S'}s_i\\
    &= \frac{(m+2)(m+1)}{2}\beta+ \sum_{i\in S'}s_i,\\
    H(\boldsymbol{\alpha}') &= l_1 + \sum_{i\in M\setminus S'} l_{i^2+1} + \sum_{i\in S'} l_{i(i+1)+1}=(m+1)\beta+m\beta+\dots+\beta + \sum_{i\in M\setminus S'}s_i\\
    &= \frac{(m+2)(m+1)}{2}\beta+ \sum_{i\in M\setminus S'}s_i.
\end{align*}
From~\eqref{proof:ass-part}, we have
\begin{align*}
    &W(\boldsymbol{\alpha}') = \frac{(m+2)(m+1)}{2}\beta+ \frac{1}{2}\beta = b
    \qquad\text{and}\qquad
    H(\boldsymbol{\alpha}') = \frac{(m+2)(m+1)}{2}\beta+ \frac{1}{2}\beta = \lambda.
\end{align*}
Thus, the answer to the D-SIPP is also `yes'.

On the other hand, if the answer to the D-SIPP is `yes',
from Corollary~\ref{cly:opt-RCL}, there must exists an RC sequence $\boldsymbol{\alpha}''=(\alpha''_1,\alpha''_2,\dots,\alpha''_u)$ with 
\begin{align}\label{proof:ass-sipp}
    W(\boldsymbol{\alpha}'')\le b \qquad\text{and}\qquad H(\boldsymbol{\alpha}'')\le \lambda.
\end{align}
Because the RC layout corresponding to $\boldsymbol{\alpha}''$ is a sorted layout, the value of $W(\boldsymbol{\alpha}'')+H(\boldsymbol{\alpha}'')-l_1$ is equal to the total side lengths of all $u+1$ bottleneck squares $v_1,v_2,\dots,v_{u+1}$, that is,
\begin{align}\label{proof:bottle}
    W(\boldsymbol{\alpha}'')+H(\boldsymbol{\alpha}'')-l_1=\sum_{j=1}^{u+1} l_{v_j}.
\end{align}
From Lemma~\ref{lem:bottleneck} and the assumption in~\eqref{assp:sprted}, we have
\begin{align}\label{proof:bottle-bound}
    &l_{v_j}\ge l_{\omega(j)}&\forall j=1,2,\dots,u+1.
\end{align}
Combining~\eqref{proof:bottle} and~\eqref{proof:bottle-bound}, we obtain
\begin{align}\label{proof:mid}
    W(\boldsymbol{\alpha}'')+H(\boldsymbol{\alpha}'')\ge l_1+\sum_{j=1}^{u+1} l_{\omega(j)}.
\end{align}
Because $n=(m+1)^2$ squares need to be packed, the size of $\boldsymbol{\alpha}''$ has to be longer than $2m$, that is, $u\ge 2m$.
From $u\ge 2m$, Lemma~\ref{lem:bottleneck} and~\eqref{eq:bottleneck}, \eqref{proof:mid} becomes
\begin{align}
\nonumber
    W(\boldsymbol{\alpha}'')+H(\boldsymbol{\alpha}'')&\ge
        l_1+\sum_{j=1}^{u+1} l_{\omega(j)}\ge
        l_1+\sum_{j=1}^{2m+1} l_{\omega(j)}= 2l_1+\sum_{i=1}^{m}l_{i(i+1)+1}+\sum_{i=1}^{m}l_{i^2+1}\\
\nonumber
    &=2(m+1)\beta + \sum_{i=1}^{m}((m-i+1)\beta)+\sum_{i=1}^{m}((m-i+1)\beta+s_i)\\
\nonumber
    &=2\beta\sum_{i=0}^{m}(m-i+1) + \sum_{i=1}^{m}s_i=(m+1)(m+2)\beta+\beta\\
\label{proof:bl-sum}
    &=b+\lambda.                   
\end{align}
Combining \eqref{proof:ass-sipp} and \eqref{proof:bl-sum}, we have
\begin{align}\label{proof:H}
&W(\boldsymbol{\alpha}'')= b=\frac{(m+1)(m+2)}{2}\beta+\frac{1}{2}\beta \text{\ \ and\ \ } H(\boldsymbol{\alpha}'')= \lambda=\frac{(m+1)(m+2)}{2}\beta+\frac{1}{2}\beta,
\end{align}
which further implies that the inequalities in~\eqref{proof:bl-sum} can be restricted to equations, that is,
\begin{align}\label{proof:eq}
    W(\boldsymbol{\alpha}'')+H(\boldsymbol{\alpha}'')=l_1+\sum_{j=1}^{u+1} l_{\omega(j)}=l_1+\sum_{j=1}^{2m+1} l_{\omega(j)}.
\end{align}
Because $l_j>0$ holds for each $j$, the second equation in~\eqref{proof:eq} implies that $u=2m$.
Following from the fact that $u=2m$ and $(m+1)^2$ squares exist, $\boldsymbol{\alpha}''$ must contain exactly $m$ AC operations and $m$ AR operations.
The first equation in~\eqref{proof:eq} implies that the side length of the $j$th bottleneck square must be $l_{\omega(j)}$.
That is, for $i=1,2,\dots, m$, the side length of the ($2i$)th bottleneck square is $(m-i+1)\beta+s_i$,
and the side length of the $(2i+1)$th bottleneck square is $(m-i+1)\beta$.
Because the first column contains a subset of these bottleneck squares, the height $H(\boldsymbol{\alpha}'')$ can be expressed as
\begin{align}\label{proof:express}
    H(\boldsymbol{\alpha}'') = C\beta + \sum_{i\in S''}s_i,
\end{align}
where $C$ is a non-negative integer and $S''$ is a subset of $M$.
Because it holds that $0\le \sum_{i\in S''}s_i\le \sum_{i\in M}s_i=\beta$,
from \eqref{proof:H} and \eqref{proof:express}, we obtain
\begin{align*}
    C=\frac{(m+1)(m+2)}{2}\qquad\text{and}\qquad \sum_{i\in S''}s_i = \frac{1}{2}\beta,
\end{align*}
which implies that $S''$ is a solution to the PP.
\end{proof}

\section{Mathematical formulations}\label{sec:formulation}
Because the SIPP is $\mathcal{NP}$-hard, no efficient algorithm exists that can solve all instances optimally in polynomial time unless $\mathcal{P}= \mathcal{NP}$.
In this section, we propose three mathematical formulations for the SIPP, and we then evaluate them by computational experiments.

The first formulation is a mixed integer linear programming (MILP) model based on the idea of the natural solution representation.
We use binary variable $x_{ijk}$ to indicate if square $k$ is placed in the $i$th row of the $j$th column.
We also define variable $y_{i}$ representing the height of row $i$, and $z_{j}$ representing the width of column $j$.
Then, the \emph{basic formulation} for the SIPP is given as 
\begin{align}
\label{mod-basic:obj}
\text{(basic formulation)}&&\min\quad&\sum_{i\in N} y_i\\
\label{mod-basic:con-ass}
&&\text{s.t.}\quad&\sum_{i\in N}\sum_{j\in N}x_{ijk}=1 &\forall k\in N\\
\label{mod-basic:con-fill}
&&& \sum_{k\in N}x_{ijk}\le 1 &\forall i\in N,\forall j\in N\\
\label{mod-basic:con-row}
&&& y_i \ge \sum_{k\in N}l_{k}x_{ijk} &\forall i\in N, \forall j\in N\\
\label{mod-basic:con-col}
&&& z_j \ge \sum_{k\in N}l_{k}x_{ijk} &\forall i\in N, \forall j\in N\\
\label{mod-basic:con-width}
&&& \sum_{j\in N} z_j\le b\\
\label{mod-basic:con-01}
&&& x_{ijk}\in \{0,1\} &\forall i\in N,\forall j\in N,\forall k\in N.
\end{align}
The objective function~\eqref{mod-basic:obj} minimizes the total height.
Constraints~\eqref{mod-basic:con-ass} ensure that every square are placed, and constraints~\eqref{mod-basic:con-fill} implies that each cell can accept at most one square.
Constraints~\eqref{mod-basic:con-row} indicate that the height of each row should be greater than or equal to the maximum side length of the squares it contains, and constraints~\eqref{mod-basic:con-col} set the width of each column.
Constraint~\eqref{mod-basic:con-width} restricts that the width of the layout should not exceed the strip width.

The second formulation we considered is based on the observation in Corollary~\ref{cly:opt-SL}, in which only binary $x_{ijk}$ is used.
The resulting formulation, which we call the \emph{sorted formulation}, becomes an integer linear programming (ILP) model as follows.
\begin{align}
\label{mod-sorted:obj}
\text{(sorted formulation)}&&\min\quad&\sum_{i\in N}\sum_{k\in N} l_kx_{i1k}\\
&&\text{s.t.}\quad&\text{\eqref{mod-basic:con-ass},\eqref{mod-basic:con-fill} and \eqref{mod-basic:con-01}}\\
\label{mod-sorted:con-width}
&&& \sum_{j\in N}\sum_{k\in N} l_kx_{1jk}\le b\\
\label{mod-sorted:con-rsorted}
&&& \sum_{k\in N}l_{k}x_{i,j-1,k} \ge \sum_{k\in N}l_{k}x_{ijk} &\forall i\in N,\forall j\in N\setminus\{1\}\\
\label{mod-sorted:con-csorted}
&&& \sum_{k\in N}l_{k}x_{i-1,j,k} \ge \sum_{k\in N}l_{k}x_{ijk} &\forall j\in N,\forall i\in N\setminus\{1\}.
\end{align}
According to Corollary~\ref{cly:opt-SL}, the height and width of the layout can be represented as 
\begin{align*}
    \sum_{i\in N}\sum_{k\in N} l_kx_{i1k}\qquad\text{and}\qquad \sum_{j\in N}\sum_{k\in N} l_kx_{1jk},
\end{align*}
respectively.
The objective function~\eqref{mod-basic:obj} and constraint~\eqref{mod-basic:con-width} in the previous formulation can be written as~\eqref{mod-sorted:obj} and \eqref{mod-sorted:con-width}.
Constraints~\eqref{mod-sorted:con-rsorted} and~\eqref{mod-sorted:con-csorted} guarantee the property that the squares in each row and column are sorted.

Next, we introduce a formulation based on the idea in Corollary~\ref{cly:opt-RCL}.
We use binary variables $\mu_i$ and $\nu_i$ to indicate whether square $i$ is in the first column and the first row, respectively.
Because the formulation focuses on RC layouts, we call it as the \emph{RC formulation}, which is given as follows.
\begin{align}
\label{mod-RC:obj}
\text{(RC formulation)}&&\min\quad&\sum_{i\in N}l_i\mu_i\\
\label{mod-RC:con-width}
&&\text{s.t.}\quad& \sum_{i\in N}l_i\nu_i\le b\\
\label{mod-RC:con-base}
&&& \mu_1=\nu_1=1\\
\label{mod-RC:con-RC}
&&&i-\left(\sum_{j=1}^{i-1}\mu_j\right)\left(\sum_{j=1}^{i-1}\nu_j\right)\le \mu_i+\nu_i\le 1  &\forall i\in N\setminus \{1\}\\
\label{mod-RC:con-01}
&&& \mu_i,\nu_i\in \{0,1\} &\forall i\in N.
\end{align}
The objective function~\eqref{mod-RC:obj} minimizes the height, that is, the total side lengths of the squares in the first column.
Constraint~\eqref{mod-RC:con-width} ensures that the width of the layout does not exceed the width of the strip.
Constraint~\eqref{mod-RC:con-base} indicates that square $1$ is placed at the bottom left.
Let $T_i$ be the set of all the bottleneck squares contained from square $1$ to $i-1$.
In $T_i$, the number of bottleneck squares in the first column and in the first row can be expressed as
$\sum_{j=1}^{i-1}\mu_j$ and $\sum_{j=1}^{i-1}\nu_j$, respectively.
Constraints~\eqref{mod-RC:con-RC} show that for square $i$ ($\ne 1$), 
if $i\le \left(\sum_{j=1}^{i-1}\mu_j\right)\left(\sum_{j=1}^{i-1}\nu_j\right)$, then the square $i$ can be placed on the space dominated by the bottleneck squares in $T_i$;
Otherwise, it has to be a new bottleneck square.

Constraints~\eqref{mod-RC:con-RC} indicates that the RC formulation is an integer quadratically constrained programming (IQCP) model.
The RC formulation contains only $2n$ of binary variables, which is much smaller than the number of $n^3$ in the other two formulations.

From assumption~\eqref{assp:sprted},
an alternative formulation with constraints~\eqref{mod-RC:con-RC} replaced by 
\begin{align*}
&&&i-\left(\sum_{j=1}^{i-1}\mu_j\right)\left(\sum_{j=1}^{i-1}\nu_j\right)\le \mu_i+\nu_i  &\forall i\in N\setminus \{1\}
\end{align*}
can also obtain the optimal value.

\subsection{Computational experiments}\label{sec:results}
All experiments were carried out on a MacBook Pro with Apple M1 Max (3.20~GHz) CPU and 64~GB memory, with computations always conducted on a single thread.
We used Gurobi Optimizer version 9.5.0 to solve the MILP, ILP and IQCP formulations.
All computations were assigned a time limit of 60 seconds per instance.

To the best of our knowledge, this is the first study on the SIPP.
We generated instances with number of squares $n\in \{10,15,20,25,30,35\}$.
For each $n$, we generated 10 instances, in which 
side length $l_j$ is an integer randomly taken from a uniform distribution over $[1,20]$,
and strip width $b$ is given as
\begin{align*}
    b = \ceil*{\left(\sum_{i\in N}l^2_i\right)^{0.5}}.
\end{align*}
Thus, we obtained 60 instances in total.

\begin{table}[ht]
\caption{Results for the three formulations}\label{tbl:res-mod}\centering
\begin{tabular}{rrrrrrrrrrrrr}
\hline
\multicolumn{1}{c}{}    & \multicolumn{1}{c}{} & \multicolumn{3}{c}{basic}                                                      & \multicolumn{1}{c}{} & \multicolumn{3}{c}{sorted}                                                     & \multicolumn{1}{c}{} & \multicolumn{3}{c}{RC}                                                         \\ \cline{3-5} \cline{7-9} \cline{11-13} 
\multicolumn{1}{c}{$n$} & \multicolumn{1}{c}{} & \multicolumn{1}{c}{\#opt} & \multicolumn{1}{c}{time} & \multicolumn{1}{c}{gap} & \multicolumn{1}{c}{} & \multicolumn{1}{c}{\#opt} & \multicolumn{1}{c}{time} & \multicolumn{1}{c}{gap} & \multicolumn{1}{c}{} & \multicolumn{1}{c}{\#opt} & \multicolumn{1}{c}{time} & \multicolumn{1}{c}{gap} \\ \hline
10                      &                      & 10                        & 0.97                     & ---                     &                      & 10                        & 0.41                     & ---                     &                      & 10                        & 0.01                     & ---                     \\
15                      &                      & 9                         & 20.64                    & 29.3\%                  &                      & 10                        & 3.09                     & ---                     &                      & 10                        & 0.17                     & ---                     \\
20                      &                      & 0                         & ---                      & 47.2\%                  &                      & 10                        & 19.35                    & ---                     &                      & 10                        & 0.78                     & ---                     \\
25                      &                      & 0                         & ---                      & 65.9\%                  &                      & 0                         & ---                      & 22.7\%                  &                      & 10                        & 5.12                     & ---                     \\
30                      &                      & 0                         & ---                      & 71.2\%                  &                      & 0                         & ---                      & 30.5\%                  &                      & 7                         & 44.58                    & 7.6\%                   \\
35                      &                      & 0                         & ---                      & 73.3\%                  &                      & 0                         & ---                      & 41.1\%                  &                      & 0                         & ---                      & 11.8\%                  \\ \hline
\end{tabular}
\end{table}

Table~\ref{tbl:res-mod} shows the results of the three formulations.
The results for each formulation show
the number of instances that solved to optimality within the time limit (``\#opt''),
the average CPU time in seconds required to achieve the optimality (``time''),
and the average optimality gap for the instances that failed to be solved to optimality within the time limit (``gap'').
The notation ``---'' in column ``gap'' indicates that the formulation exactly solved all the 10 instances within 60 seconds,
and the same notation in column ``time'' indicates that the formulation was not able to exactly solve any of the 10 instances.

From Table~\ref{tbl:res-mod}, we observe that all the three formulations can solve instances with $n=10$.
The basic formulation failed to solve all the instance with $n\ge 20$, and it obtained solutions with large optimality gaps for all these instances.
The sorted formulation performed better than the basic formulation, obtaining optimal solutions for all the instances with $n\le 20$ in 60 seconds.
Benefiting from RC layout representation, the RC formulation outperforms the other two formulations.
It exactly solved all the instances with $n\le 25$ in a shorter average time, and it solved 7 out of 10 instances with $n=30$.
For the instances with $n=35$, all the formulations failed to exactly solve any instances within 60 seconds,
but the RC formulation obtained solutions with a smaller average gap.

From the above results, we observe that the RC formulation outperforms the other two formulations.
The RC formulation exactly solved all the instances with $n\le 25$ in a short time (less than 10 seconds).
However, it failed to exactly solve any instances with $n\ge 35$ within 60 seconds.

\section{Fully-polynomial approximation scheme}\label{sec:fptas}
Corollary~\ref{cly:opt-RCL} indicates that to solve the SIPP, we can design an approach to search for the best solution among all RC layouts.
In this section, we propose a \emph{dynamic programming} (DP) method running in pseudo-polynomial time based on Corollary~\ref{cly:opt-RCL},
and then we extend the DP to a \emph{fully polynomial approximation scheme} (FPTAS).

\subsection{Dynamic programming}\label{sec:dp}
Consider a subproblem in which the width of the strip is $k$ and the squares $1,2,\dots,ij$ should be independently packed to a layout of $i$ rows and $j$ columns.
We use $f(i,j,k)$ to denote the minimum height of the strip for this subproblem.
We show that $f(i,j,k)$ can be computed recursively with $i+j\ge 3$.
Let $\boldsymbol{\alpha}^*=(\alpha^*_1,\alpha^*_2,\dots,\alpha^*_{i+j-2})$ be an optimal RC sequence to the subproblem.
First, we consider the case when $\alpha^*_{i+j-2}=\mathrm{AR}$.
In this case, the sequence $(\alpha^*_1,\alpha^*_2,\dots,\alpha^*_{i+j-3})$ must be an optimal solution for the subproblem corresponding to $f(i-1,j,k)$;
Otherwise, using an optimal solution to the subproblem corresponding to $f(i-1,j,k)$ with an AR operation will lead to a solution with smaller height than $\boldsymbol{\alpha}^*$.
Because the $i$th row has a height of $l_{ij-j+1}$, the optimal value for the case when $\alpha^*_{i+j-2}=\mathrm{AR}$ should be $f(i-1,j,k)+l_{ij-j+1}$.
Similarly, if $\alpha^*_{i+j-2}=\mathrm{AC}$, then the $j$th column has a width of $l_{ij-i+1}$,
and $(\alpha^*_1,\alpha^*_2,\dots,\alpha^*_{i+j-3})$ must be an optimal solution for the subproblem corresponding to $f(i,j-1,k-l_{ij-i+1})$.
The above analysis of the two cases shows that the smaller value between $f(i-1,j,k)+l_{ij-j+1}$ and $f(i,j-1,k-l_{ij-i+1})$ is the optimal value to the subproblem corresponding to $f(i,j,k)$.
Thus, for each $i\in N$, $j\in N$ and $k\in \{l_1,l_1+1,\dots,b\}$, we compute $f(i,j,k)$ using the recurrence formula
\begin{align}\label{dp:rec}
    f(i,j,k)=\min\{f(i-1,j,k)+l_{ij-j+1},f(i,j-1,k-l_{ij-i+1})\},
\end{align}
where the boundary condition is given as follows:
\begin{align}\label{dp:bd}
f(i,j,k)=
    \begin{cases}
   l_1 &\text{if $i=j=1$ and $l_1\le k\le b$}\\
   +\infty &\text{if $i\le 0$ or $j\le 0$ or $k< l_1$}.
    \end{cases}
\end{align}
The boundary condition implies that the base layout has a height of value $l_1$.

The optimal value to the SIPP is
\begin{align}\label{dp-opt-naive}
\min_{i\in N,j\in N} \{f(i,j,b)\mid ij\ge n\}.
\end{align}
This optimal value can be obtained in $\mathcal{O}(n^2b)$ time by computing all necessary values of $f(i,j,k)$ for $i$ from $1$ to $n$, $j$ from $1$ to $n$, and $k$ from $l_1$ to $b$.
However, we observe that $(i,j)$ pairs included in \eqref{dp-opt-naive} may lead to a solution that contains empty rows and columns, that is, all the squares in that row or column are dummy squares.
Based on this observation, the optimal candidates can be further restricted to $(i,j)$ pairs that satisfy the conditions $i(j-1)<n$ and $(i-1)j<n$,
which implies that the optimal value can be obtained by
\begin{align*}
\min_{i\in N,j\in N} \{f(i,j,b)\mid ij\ge n, i(j-1)<n, (i-1)j<n\}.
\end{align*}
All these restricted candidates can be computed more efficiently by an algorithm shown in Algorithm~\ref{algo:dp}.
The total running time is improved to $\mathcal{O}(n^{1.5}b)$.
\begin{algorithm}[ht]
    \caption{Dynamic programming for the SIPP}\label{algo:dp}
    $z^*\gets +\infty$\;
    \lFor{$k=l_1,l_1+1,\dots,b$}{$f(1,1,k)\gets l_1$}
    \For{$i=1,2,\dots,\ceil*{n^{0.5}}$}{
        \For{$j=1,2,\dots,\ceil*{\frac{n}{i}}$}{
            \For{$k=l_1,l_1+1,\dots,b$}{
                Compute $f(i,j,k)$ according to \eqref{dp:rec}--\eqref{dp:bd}\;
           }
       }
        \lIf{$z^*>f(i,\ceil*{\frac{n}{i}},b)$}{$z^*\gets f(i,\ceil*{\frac{n}{i}},b)$}
   }
    \For{$j=1,2,\dots,\ceil*{n^{0.5}}$}{
        \For{$i=\ceil*{n^{0.5}}+1,\ceil*{n^{0.5}}+2,\dots,\ceil*{\frac{n}{j}}$}{
            \For{$k=l_1,l_1+1,\dots,b$}{
                Compute $f(i,j,k)$ according to \eqref{dp:rec}--\eqref{dp:bd}\;
           }
       }
        \lIf{$z^*>f(\ceil*{\frac{n}{j}},j,b)$}{$z^*\gets f(\ceil*{\frac{n}{j}},j,b)$}
   }
    \Return $z^*$\;
\end{algorithm}

Note that the space complexity of Algorithm~\ref{algo:dp} is $\mathcal{O}(n^{1.5}b)$, which can be further reduced to $\mathcal{O}(n^{0.5}b)$ if the optimal solution is not needed to be recovered. 
Algorithm~\ref{algo:r-dp} shows the algorithm running with $\mathcal{O}(n^{1.5}b)$ space, where all values of $f(i,j,k)$ are computed using two 2-dimensional arrays $f^\mathrm{odd}$ and $f^\mathrm{odd}$ that store at most $\ceil*{n^{0.5}}\cdot b$ values.

\begin{algorithm}[ht]
    \caption{Dynamic programming for the SIPP with $\mathcal{O}(n^{0.5}b)$ space}\label{algo:r-dp}
    $z^*\gets +\infty$\;
    \lFor{$k=l_1,l_1+1,\dots,b$}{$f(1,1,k)\gets l_1$}
    \For{$i=1,2,\dots,n$}{
        \For{$j=1,2,\dots,\ceil*{n^{0.5}}$}{
            \For{$k=l_1,l_1+1,\dots,b$}{
                \uIf{$i$ is odd.}{
                    $f^\mathrm{odd}(j,k)=\min\{f^\mathrm{even}(j,k)+l_{ij-j+1},f^\mathrm{odd}(j-1,k-l_{ij-i+1})\}$\;
                    \lIf{$k=b$ and $j=\ceil*{\frac{n}{i}}$ and $z^*>f^\mathrm{odd}(j,b)$}{$z^*\gets f^\mathrm{odd}(j,b)$}
                }
                \Else{
                    $f^\mathrm{even}(j,k)=\min\{f^\mathrm{odd}(j,k)+l_{ij-j+1},f^\mathrm{even}(j-1,k-l_{ij-i+1})\}$\;
                    \lIf{$k=b$ and $j=\ceil*{\frac{n}{i}}$ and $z^*>f^\mathrm{even}(j,b)$}{$z^*\gets f^\mathrm{even}(j,b)$}
                }
           }
       }
   }
    \For{$j=1,2,\dots,n$}{
        \For{$i=1,2,\dots,\ceil*{n^{0.5}}$}{
            \For{$k=l_1,l_1+1,\dots,b$}{
                \uIf{$j$ is odd.}{
                    $f^\mathrm{odd}(i,k)=\min\{f^\mathrm{odd}(i-1,k)+l_{ij-j+1},f^\mathrm{even}(i,k-l_{ij-i+1})\}$\;
                    \lIf{$k=b$ and $i=\ceil*{\frac{n}{j}}$ and $z^*>f^\mathrm{odd}(i,b)$}{$z^*\gets f^\mathrm{odd}(i,b)$}
                }
                \Else{
                    $f^\mathrm{even}(i,k)=\min\{f^\mathrm{even}(i-1,k)+l_{ij-j+1},f^\mathrm{odd}(i,k-l_{ij-i+1})\}$\;
                    \lIf{$k=b$ and $i=\ceil*{\frac{n}{j}}$ and $z^*>f^\mathrm{even}(i,b)$}{$z^*\gets f^\mathrm{even}(i,b)$}
                }
           }
       }
   }
    \Return $z^*$\;
\end{algorithm}
\subsection{Design of FPTAS}
For some combinatorial optimization problems, a DP method can be converted to an FPTAS \cite{woeginger2000does}.
However, we cannot use a common scaling technique directly for the SIPP, because scaling on $l_j$ not only influences the objective value, but also may change the feasibility of the original problem.
More specifically, let $t$ ($\ge 1$) be a scaling factor,
because the DP method requires integer inputs, 
we have to round up by $\bar{l}_j\gets \ceil*{\frac{l_j}{t}}$ for each $j\in M$ and $\bar{b}\gets \ceil*{\frac{b}{t}}$ to obtain a new instance.
But, there is no containment or inclusion relationship between the feasible regions of the original instance and the new instance.
For example, consider an instance with 2 squares with $l_1=15$, $l_2=14$ and $b=29$.
The scaling factor $t=10$ leads to a new instance with $\bar{l}_1=2$, $\bar{l}_2=2$, $\bar{b}=3$, and thus the layout with square 1 and 2 in the same row becomes an infeasible solution for the new instance, which is feasible to the original instance.
On the other hand, if we consider an instance with $l_1=10$, $l_2=10$ and $b=19$.
The same scaling factor $t=10$ leads to a new instance with $\bar{l}_1=1$, $\bar{l}_2=1$, $\bar{b}=2$.
This time, the layout with square 1 and 2 in the same row is feasible to the new instance but infeasible to the original one.

To overcome this difficulty, before we design the algorithm we consider an generalization of the SIPP, which we call the \textsc{Rectangular Independent Packing Problem} (RIPP).
In the RIPP, instead of $n$ squares we are given $n$ rectangles, each rectangle $i$ with a width $w_j$ and a height $h_j$.
In this paper, we refer as \mbox{RIPP-C} the RIPP satisfying conditions
\begin{align}\label{ass:esipp}
    w_1\ge w_2\ge\dots \ge w_n \qquad\text{and}\qquad h_1\ge h_2\ge\dots \ge h_n.
\end{align}
Note that, although the RIPP-C is a generalization of the SIPP, Corollary~\ref{cly:opt-SL} and Corollary~\ref{cly:opt-RCL} hold for RIPP-C as well.

We use the RIPP-C to design an FPATS for the SIPP.
The DP method described in Section~\ref{sec:dp} can be used to solve the RIPP-C, and the running time is still $\mathcal{O}(n^{1.5}b)$.
In this section, we introduce a similar DP for the RIPP-C to design an FPTAS.
Let $g(i,j,k)$ denote the minimum width of the layout when rectangles $1,2,\dots,ij$ should be independently packed to a layout of $i$ rows and $j$ columns and the height of layout can not exceed $k$.

By setting the the boundary conditions as 
\begin{align}\label{dp:g-bound}
g(i,j,k)=
    \begin{cases}
   w_1 &\text{if $i=j=1$ and $k\ge h_1$}\\
   +\infty &\text{if $i\le 0$ or $j\le 0$ or $k< h_1$,}
    \end{cases}
\end{align}
each $g(i,j,k)$ can be computed by the following formula
\begin{align}\label{dp:g-recur}
    g(i,j,k)=\min\{g(i,j-1,k)+w_{ij-i+1},g(i-1,j,k-h_{ij-j+1})\}.
\end{align}

Let $H_{\Sigma}=\sum_{i\in N}h_i$ denote the total heights of all rectangles.
The optimal value to the \mbox{RIPP-C} is
\begin{align}\label{dp:g-opt}
\min_{\substack{i\in N,j\in N,\\k=h_1,h_1+1,\dots,H_{\Sigma}}} \{k\mid g(i,j,k)\le b, ij\ge n, i(j-1)<n, (i-1)j<n\}.
\end{align}
Using a similar algorithm as in Algorithm~\ref{algo:dp}, the running time for computing the optimal value is $\mathcal{O}(n^{1.5}H_{\Sigma})$.

Next, we come back to describe the FPTAS for the SIPP.
Let $\epsilon$ be an arbitrary small constant.
By introducing a scaling factor
\begin{align}\label{def:t}
    t= \max\left\{\frac{\epsilon l_1}{n},1\right\},
\end{align}
we transform an SIPP instance $(\boldsymbol{l},b)$ to an RIPP-C instance $(\boldsymbol{\bar{w}},\boldsymbol{\bar{h}},\bar{b})$ as follows:
\begin{align*}
    &\bar{w}_i = l_i, \bar{h}_i =\ceil*{\frac{l_i}{t}}\qquad \forall i\in N;\qquad\qquad\bar{b} = b.
\end{align*}
Let $\bar{\boldsymbol{A}}$ be the RC layout for the RIPP-C obtained by the DP described in~\eqref{dp:g-bound}--\eqref{dp:g-opt}.
\begin{lemma}
For the original SIPP, the relative error of solution $\bar{\boldsymbol{A}}$ is bounded by $\epsilon$.
\end{lemma}
\begin{proof}
Assume that $\boldsymbol{A}^*$ is an optimal RC layout to the original SIPP instance.
Layouts $\boldsymbol{A}^*$ and $\bar{\boldsymbol{A}}$ are feasible for both instances,
because $\bar{b} = b$ and $\bar{w}_i = l_i$ for each $i$.
For convenience, in this proof we use $H(\boldsymbol{A})$ and $\bar{H}(\boldsymbol{A})$ to denote the height of a layout $\boldsymbol{A}$ for the SIPP instance $(\boldsymbol{l},b)$ and the \mbox{RIPP-C} instance $(\boldsymbol{\bar{w}},\boldsymbol{\bar{h}},\bar{b})$, respectively.

Because $\bar{H}({\boldsymbol{A}})=H(\boldsymbol{A}^*)$ can be easily derived when $t=1$ in~\eqref{def:t}, we focus on the case when $t=\epsilon l_1/n$.
Layout $\bar{\boldsymbol{A}}$ is an optimal solution for the RIPP-C, and thus we have $\bar{H}(\bar{\boldsymbol{A}}) \le \bar{H}(\boldsymbol{A}^*)$.
Because $\bar{h}_i =\ceil*{\frac{l_i }{t}}$, $l_i\le t\bar{h}_i\le t+l_i$ holds for each $i$ in $N$.
For layouts $\bar{\boldsymbol{A}}$ and $\boldsymbol{A}^*$,
we use $\bar{S}$ and $S^*$ to denote the sets that consist of all the items in the first column, respectively.
Then, we have
\begin{align*}
H(\bar{\boldsymbol{A}}) 
&= \sum_{i\in \bar{S}} l_i 
\le \sum_{i\in \bar{S}} t\bar{h}_i 
= t\sum_{i\in \bar{S}} \bar{h}_i = t\bar{H}(\bar{\boldsymbol{A}})\le t\bar{H}(\boldsymbol{A}^*) = t\sum_{i\in S^*} \bar{h}_i 
 \\
&= \sum_{i\in S^*} t\bar{h}_i\le \sum_{i\in S^*} (t+l_i)
\le nt+\sum_{i\in S^*} l_i= \epsilon l_1 + H(\boldsymbol{A}^*)\\
&\le (1+\epsilon)H(\boldsymbol{A}^*).
\end{align*}
\end{proof}

The time complexity of the FPTAS is dominated by the running time of the DP in~\eqref{dp:g-bound}--\eqref{dp:g-opt} for the RIPP-C instance $(\boldsymbol{\bar{w}},\boldsymbol{\bar{h}},\bar{b})$,
that is,
\begin{align*}
    O\left(n^{1.5}\left(\sum_{i\in N}\bar{h}_i\right)\right)&=O\left(n^{1.5}\left(\sum_{i\in N}\frac{l_i}{t}\right)\right)=O\left(n^{1.5}\left(\frac{nl_1}{t}\right)\right)=O\left(\frac{n^{3.5}}{\epsilon}\right),
\end{align*}
which is polynomially bounded by both the input size $n$ and $1/\epsilon$.

\section{Extensions}\label{sec:extension}
In Section~\ref{sec:fptas}, we considered the RIPP-C as an extention of the SIPP.
In this section, we introduce several other useful extensions that may relate to practical cases.

\subsubsection*{SIPP with positive-thickness partitions}
Until now, we have assumed that the thickness of partitions is zero.
However, many real-world applications may include partitions of $\eta$ ($\ge 0$) thickness.
To deal with thickness $\eta$,
we can add $\eta$ to each $l_i$ and $b$ in advance,
and solve the corresponding problem without thickness.
After such a preprocessing, the optimal value of the problem becomes a constant value $\eta$ larger than the optimal value of the original problem.

Accordingly, all the results obtained in this paper are applicable to a RIPP-C where a horizontal partition and/or a vertical partition can have different thicknesses.

\subsubsection*{SIPP with only horizontal partitions}
We also consider a relaxation of the SIPP, where only horizontal partitions have to be used.
That is, several squares can be placed horizontally in the same row so that they do not overlap and the total width does not exceed $b$.
This relaxation problem becomes a special case of the \textsc{2-dimensional Level Strip Packing Problem} (2LSPP) when all the items are given as squares.
The 2LSPP is known to be $\mathcal{NP}$-hard in the strong sense, because it contains the bin-packing problem
as a special case~\cite{lodi2002two}.
We show the SIPP with only horizontal partitions (SIPP-H) is also $\mathcal{NP}$-hard in the strong sense.
\begin{theorem}
\label{thm:sipp-h-hard}
    The SIPP-H is $\mathcal{NP}$-hard in the strong sense.
\end{theorem}
\begin{proof}
    We show that the \textsc{3-partition problem} (3PP)~\cite{garey1979computers}, which is known to be $\mathcal{NP}$-complete in the strong sense, is reducible to the SIPP-H.
    
    We are given in the 3PP a set $M = \{1, 2, \dots, 3c\}$, each element $i$ with a positive value $q_j$. Let $B = \frac{1}{c}\sum_{j\in M} q_j$ and each $q_j$ satisfies $B/4 < q_j < B/2$.
    The 3PP determines whether there exists a partition $M_1, M_2, \dots, M_c$ of $M$ such that for each $M_i$ ($i=1,2,\dots,c$), $\sum_{j \in M_i} q_j = B$ holds. 
    
    We transfer an arbitrary instance of the 3PP to an instance of the SIPP-H with $n = 4c$ squares as follows:
    \begin{align*}
        b = 2B+1;\qquad\qquad
        &l_i=
        \begin{cases}
        B+1&\text{if $i=1,2,\dots,c$}\\
        q_{i-c}&\text{if $i=2c,2c+1,\dots,4c$.}
        \end{cases}
    \end{align*}
    Because squares $1,2,\dots,c$ must be placed in different rows due to the strip width limitation,
    $\sum_{j=1}^c l_i = c(B+1)$ is a valid lower bound of the SIPP-H.
    For a row in which one of the squares $1,2,\dots,c$ is placed, the remaining width is $B$.
    Thus, the SIPP-H has a solution whose height equals to the lower bound $c(B+1)$ if and only if the answer of the 3PP is `yes.'
\end{proof}
Thus, different from the original SIPP,
the SIPP-H does not admit a pseudo-polynomial time algorithm unless $\mathcal{P}=\mathcal{NP}$.

\subsubsection*{SIPP with only vertical partitions}
Similarly, the SIPP with only vertical partitions is also $\mathcal{NP}$-hard in the strong sense.
\begin{theorem}
    The SIPP with only vertical partitions is $\mathcal{NP}$-hard in the strong sense.
\end{theorem}
By a proof similar to that of Theorem~\ref{thm:sipp-h-hard}, we can show that the SIPP without using horizontal partitions is reducible from the 3-partition problem.

\subsubsection*{$k$-dimensional SIPP}
We consider extending the SIPP to a 3-dimensional version (3SIPP). 
More specifically, given a strip with width $b$ and length $\lambda$, and a set of cubes, each cube $i$ with side length $l_i$, the 3SIPP aims to independently pack all cubes by using division plates perpendicular to the $x$, $y$ or $z$ axis to minimize the height of the strip.
Such an extension has many real-world applications, such as packing fragile spherical objects in transportation, or cutting out spherical or cube objects from materials.

For the 3SIPP, all the results in Sections~\ref{sec:solution-represntation}, \ref{sec:formulation} and \ref{sec:dp} can be extended.
Thus, we can design a dynamic programming algorithm with a time complexity of $O(n^{\frac{11}{6}}b\lambda)$.

Theoretically, the SIPP can be extended to a $k$-dimensional version ($k$SIPP), where the capacity $b_i$ of each dimension $i$ is given for $i=1,2,\dots,k-1$.
For the $k$SIPP, we can also design a dynamic programming extended from Sections~\ref{sec:dp} with a running time of $O(kn^{h(k)}\prod_{i=1}^{k-1}b_i)$ where $h(k)=\sum_{i=1}^k \frac{1}{i}$.

However, for a $k$SIPP with $k\ge 3$, we show that there is no FPTAS unless $\mathcal{P}=\mathcal{NP}$ even with $k=3$.
\begin{theorem}
   The 3SIPP does not admit an FPTAS unless $\mathcal{P}=\mathcal{NP}$.
\end{theorem}
\begin{proof}
We reduce the PP with $m$ elements to the 3SIPP with $(m+1)^2$ cubes and a strip whose width and length are $b$ and $\lambda$ respectively.
The transformation is the same as in \eqref{proof:transfer-start}--\eqref{proof:transfer-end}.
From the proof of Theorem~\ref{thm:sipp-complexity}, we know that the optimal value (the minimum height of the strip) to the corresponding 3SIPP is $l_1=(m+1)\beta$ if the PP has a solution.
Otherwise, the optimal value must be at least $l_1+l_{(m+1)^2}=(m+2)\beta$.
Thus, an FPTAS for the 3SIPP can be used to distinguish between these two cases in polynomial time in the input size $m$ of the PP.
\end{proof} 
\section{Conclusion}\label{sec:conclusion}
In this study, we considered the the square independent packing problem (SIPP) that is a two-dimensional square strip packing problem with a separation constraint.
We showed that the SIPP is $\mathcal{NP}$-hard, and we designed three solution representations including a compact representation that we call the row-column (RC) sequence representation.
Three mathematical formulations were then proposed based on the different solution representations.
By experimental evaluation, one of the formulations based on the RC sequence dominated the others.
Also based on the RC-sequence representation, we further proposed an FPTAS for the SIPP.
Future research could be devoted to designing efficient algorithms for the extensions introduced in Section~\ref{sec:extension}.









\begin{thebibliography}{1}
\expandafter\ifx\csname url\endcsname\relax
  \def\url#1{\texttt{#1}}\fi
\expandafter\ifx\csname urlprefix\endcsname\relax\def\urlprefix{URL }\fi
\expandafter\ifx\csname href\endcsname\relax
  \def\href#1#2{#2} \def\path#1{#1}\fi

\bibitem{dyckhoff1990typology}
H.~Dyckhoff, A typology of cutting and packing problems, European Journal of
  Operational Research 44~(2) (1990) 145--159.

\bibitem{wascher2007improved}
G.~W{\"a}scher, H.~Hau{\ss}ner, H.~Schumann, An improved typology of cutting
  and packing problems, European Journal of Operational Research 183~(3) (2007)
  1109--1130.

\bibitem{junior2022rectangular}
A.~N. J{\'u}nior, E.~Silva, M.~Francescatto, C.~B. Rosa, J.~Siluk, The
  rectangular two-dimensional strip packing problem real-life practical
  constraints: A bibliometric overview, Computers \& Operations Research 137
  (2022) 105521.

\bibitem{lodi2002two}
A.~Lodi, S.~Martello, M.~Monaci, Two-dimensional packing problems: A survey,
  European Journal of Operational Research 141~(2) (2002) 241--252.

\bibitem{bezerra2020models}
V.~M. Bezerra, A.~A. Leao, J.~F. Oliveira, M.~O. Santos, Models for the
  two-dimensional level strip packing problem: A review and a computational
  evaluation, Journal of the Operational Research Society 71~(4) (2020)
  606--627.

\bibitem{bettinelli2008branch}
A.~Bettinelli, A.~Ceselli, G.~Righini, A branch-and-price algorithm for the
  two-dimensional level strip packing problem, 4OR 6~(4) (2008) 361--374.

\bibitem{lodi2004models}
A.~Lodi, S.~Martello, D.~Vigo, Models and bounds for two-dimensional level
  packing problems, Journal of Combinatorial Optimization 8 (2004) 363--379.

\bibitem{garey1979computers}
M.~R. Garey, D.~S. Johnson, Computers and Intractability: A Guide to the Theory
  of NP-Completeness, Freeman, 1979.

\bibitem{woeginger2000does}
G.~J. Woeginger, When does a dynamic programming formulation guarantee the
  existence of a fully polynomial time approximation scheme ({FPTAS})?, INFORMS
  Journal on Computing 12~(1) (2000) 57--74.

\end{thebibliography}
\end{document}